\documentclass[twocolumn,aps]{revtex4}

\usepackage{graphicx}
\newcommand{\be}{\begin{equation}}
\newcommand{\ee}{\end{equation}}
\newcommand{\ben}{\begin{eqnarray}}
\newcommand{\een}{\end{eqnarray}}

\begin{document}

\title{ Halperin-Lubensky-Ma effect in type-I
superconducting films }

\author{L.M. Abreu, C. de Calan} 
\address{{\it Centre de Physique Th{\'e}orique, Ecole Polytechnique, 91128
Palaiseau, France}}
 
\author{A.P.C. Malbouisson} 
\address{{CBPF/MCT, Rua Dr. Xavier Sigaud, 150, Rio de Janeiro RJ, Brazil}}

\begin{abstract}
\noindent In this note we employ concurrently  techniques  
of generalized $zeta$-functions and compactification methods introduced in previous 
publications, to study 
 the Halperin-Lubensky-Ma theory of induced weak first-order phase transitions  
 applied to type-I superconducting films. We obtain closed formulas to 
the critical temperature and to the size temperature as functions of the film thickness.

\noindent PACS number(s): 74.20.-z, 05.10Cc, 11.25.Hf

\noindent Keywords: superconductivity, superconducting films

\end{abstract}

\maketitle

The Halperin-Lubensky-Ma (HLM) effect appeared about three decades ago \cite{HLM}. It predicts a weak first-order phase transition in superconductors. This fact emerges by considering in the Ginzburg-Landau (GL) model the interaction between the intrinsic magnetic fluctuations and the order parameter. More generally, all the physical systems that are described by the Abelian-Higgs model present this phenomenon. Some examples are: the nematic-smectic A phase transition in liquid crystals \cite{HLM,LC}, with the interaction between the smectic scalar order parameter and the vector director;  the massless scalar electrodynamics in field theory  \cite{CW}, which presents the gauge field acquiring mass as an effect of its coupling with the scalar field.

However, the size temperature associated to the HLM effect was determined to be too small which makes it very difficult to be detected experimentally. Recently, the mentioned effect has been studied in the context of type-I superconducting films \cite{FSU,STU,ST,AM,AM2}. In particular, Refs. \cite{FSU,STU,ST} suggest the enhancement of first-order transition in superconducting films with respect to that in bulk materials, which at least qualitatively is corroborated in Ref. \cite{AM2}. 

In order to have a better understanding of HLM effect in films, in this note we extend some questions raised in Ref. \cite{AM2}. In a field theoretical approach, we consider the GL model submitted to confinement between two planes a distance $L$ apart from one another. This is done using a spatial compactification formalism presented in recent works \cite{E,A,JM,AMMS,AMR}. Physically,  for dimension $d=3$, $L$ corresponds to the thickness of a film-like superconducting material. We take into account the gauge fluctuations in absence of external magnetic field, and in the approximation of uniform order parameter $\psi \left( {\bf x}\right)= const.$  We investigate the critical behavior of the system as a
function of the film thickness $L$, and in particular we focus on the $L$-dependence of the interesting thermodynamical quantities, as well as we discuss the plausibility of the presented results.

Let us consider the Hamiltonian density of the GL model in Euclidean $d$-dimensional space, 
\begin{eqnarray}
{\cal H} &=&\frac{1}{2} \left( \partial _{\mu }A_{\nu } -\partial _{\nu }A_{\mu
} \right)^{2}+|(\partial _{\mu }-ieA_{\mu
})\psi |^{2}  \nonumber \\
&& + m_{0}^{2}|\psi |^{2}+\frac{\lambda }{2}(|\psi |^{2})^{2},  \label{eq1}
\end{eqnarray}
where $\psi =\psi \left( {\bf x}\right) $ is a complex field, and $m_{0}^{2}$
is the bare mass and $A_{\mu} = A_{\mu} \left( {\bf x}\right) $ ($\mu ,\nu =1...d$) is the gauge field.
Notice that we are working in the mean field convention \cite{K}. Accordingly, we use natural units ($\hbar=c=1$), and employ $\xi_{0}$ (the intrinsic coherence length) and $K_{B}T_{0}$ ($T_{0}$ corresponds to the bulk mean field transition temperature), as length and energy scales, respectively. The fields and coordinates are rescaled by $\psi \rightarrow \psi_{new}=\sqrt{\xi_{0}/k_{B} T_{0}}\, \psi$, $A \rightarrow A_{new}=\sqrt{\xi_{0}/k_{B} T_{0}}\, A$
and $x \rightarrow x_{new}=x/\xi_{0}$. Thus, the Hamiltonian density in Eq.(\ref{eq1}) is dimensionless. Also the coupling constants $e$ and $\lambda$ are dimensionless, as well as the bare mass, which is given by $m_{0}^{2}=T/T_{0}-1$.
 
The definition of the field as $\psi _{new} =\phi e^{i\theta }$, together with the gauge transformation ${\bf A}_{new} \rightarrow {\bf a}-1/e{\bf \nabla }\theta $, allows us to work with ${\cal H}$ manifestly gauge invariant \cite{K,N}. Then, the integrations over $\psi _{new}$, $\psi ^{\star }_{new}$ and ${\bf A}_{new}$ in the generating functional change
to integrals over $\phi ,\;{\bf a}$ and $\theta $. Since the functional integration over 
$\theta $ is
Gaussian, it can be performed directly. Taking explicitly the approximation in which the scalar field is spatially uniform, that is $\phi \left( {\bf x}\right) \approx \phi=const.$, 
the functional integral over ${\bf a}$ can be done, yielding the following free energy density
${\cal F}= {\cal F}(\phi)$,
\begin{eqnarray}
{\cal F} = \frac{1}{2}m_{0}^{2}\phi ^{2}+\frac{\lambda }{8}\phi ^{4} + V(\phi),  
\label{eq2}
\end{eqnarray}
where 
\begin{equation}
V(\phi) = \frac{1}{2V}\ln \left\{ \left[ \left( -\nabla ^{2}+e^{2}\phi ^{2}\right)\delta _{\mu \nu } +\partial _{\mu }\partial _{\nu }\right] \delta \left( 
{\bf x-y}\right) \right\} .   \label{eq3}
\end{equation}
is the contribution coming from the ${\bf a}$-integration. In the case that the model has no restrictions of spatial confinement, it can be shown that performing the trace in the right 
hand side of Eq. (\ref{eq3}) and using the momentum space representation,  
 a $\phi ^3$-dependent term is generated  
for $d=3$, which suggests that systems described by the free energy (\ref{eq2}) undergoes a first-order phase transition \cite{HLM}. 

However, for situations in which the model is restricted to boundary conditions, Eq. (\ref{eq3}) must be treated with some additional techniques. Let us consider the system confined between two parallel planes, normal
to the $x_{d}$-axis, a distance $L$ apart from one another (we will use the dimensionless form $l=L/\xi _0$). We use coordinates ${\bf x}=({\bf z},x_{d})$, where ${\bf z}$ is a $(d-1)$
-dimensional vector, with corresponding momenta ${\bf k}=({\bf q},k_{d})$.
Accordingly to \cite{E}, in this confined situation, the term corresponding to $V(\phi)$ in Eq. (\ref{eq3}) should be written as
\be
V(\phi , l) = -\frac{1}{2l} \eta'(0;\phi, l), 
\label{eq4}
\ee
where the prime means  derivation with respect to the first argument in $\eta$.
The function $\eta $ is associated with the eigenvalues of the operator given by Eq. (\ref{eq3})  
with the compactification of one dimension, 
\be
\eta(s;\phi, l)= \sum_{n=-\infty}^{\infty} \int
\frac{d^{d-1} k}{(2\pi)^{d-1}} \left[ \left(\frac{2\pi n}{l}\right)^2 + {\bf k}^2 +
  e^2 \phi^2 \right]^{-s}. 
\label{eq5}
\ee
We can perform the integration in the above equation with the help of dimensional regularization techniques; we get  
\ben
\eta(s; c, l) & =& \left(\frac{\pi}{l^2}\right)^{\frac{d-1}{2}}
\frac{\Gamma \left( s- \frac{d-1}{2} \right)}{\Gamma \left( s \right)}
 \left(\frac{l}{2\pi}\right)^{2s} \nonumber \\
& & Z_{1}^{c^2}\left( s-\frac{d-1}{2};w_1 \right),  
\label{eq6}
\een
where we have defined 
\be
Z_{N}^{c^2}\left(\nu; \{w_{i} \}
\right)=\sum_{  \{ n_i\}=- \infty }^{\infty}
\left[ w_{1}^{2}n_{1}^{2} +...+w_{N}^{2}n_{N}^{2} +c^2 \right]^{-\nu},
\label{eq7}
\ee
with $w_i=l_1/l_i\; (i=1,...,N)$ and $c^2 = \left(e \phi l/2\pi\right)^2 $. 
$N$ stands for the number of compactified dimensions, in the general case of a $N$-dimensional compactified subspace($N<d$). For us it is $N=1$, and we identify $l_1 \equiv l$. 

As is shown in Ref. \cite{E}, Eq. (\ref{eq6}) and consequently Eq. (\ref{eq4})
can be analytically extended to all values of $d$. For odd $d$(remember that we are interested in $d=3$ with one compactified dimension), the analytical continuation of $V(\phi , L)$ 
has the following expression\cite{E}, 
\ben
V(c, l)  = -\frac{1}{2l} \left(\frac{\pi}{l^2}\right)^{p}
\frac{(-1)^{p}}{p!} Z_{1}^{'\;c^{2}}\left( -p;w_1 \right),
\label{eq8}
\een
where $p= \frac{d-1}{2}$. For small values of $c^2$, $c^2<<1$, it
is possible to use the binomial expansion for $ Z_{1}^{c^{2}}$ in order to expand it in powers of the field $\phi$, 
\be
Z_{1}^{c^2}\left(q; w_1 \right) = c^{-2q}+ \sum_{j=0}^{\infty}
\frac{(-1)^{j}}{j!} T(q,j)  Z_{1}\left( q+j; w_1  \right) c^{2j}, 
\label{eq9}
\ee
where 
\be
 T(q,j)= \frac{ \Gamma (q+j)}{\Gamma (q)},
\label{eq10}
\ee
and
\be
Z_{N}\left(\nu; \{ w_i \} \right) = \sum_{\{ n_i\}=- \infty
}^{\infty} {}' 
  \left[ w_{1}^{2}n_{1}^{2} +...+w_{N}^{2}n_{N}^{2} \right]^{-\nu}.
\label{eq11}
\ee
The prime in (\ref{eq11}) means that the term $\{ n_i \} =0$ 
is excluded from the summation. For $N=1$, $w_1 =1$, and it is easily seen 
that $Z_{1}\left(\nu,w_1 = 1 \right)$ is related to the Riemann zeta-function, $\zeta
(\nu)$ by,  
\be 
Z_{1}\left(\nu,w_1 \right)=2 \zeta (2 \nu), 
\label{eq12}
\ee
Then the derivative of $Z_{1}^{c^2}\left(q; w_1 = 1 \right)$ 
for $c^2<<1$ 
in Eq. (\ref{eq9}) above with respect to $q$ gives the following expression,
\ben
Z_{1}^{'\;c^2<<1}\left(q; w_1 =1 \right)  = -2c^{-2q}\ln{c} + 2\sum_{j=0}^{\infty}
c^{2j}\frac{(-1)^j}{j!} \nonumber \\
\times \left\{ T'(q,j) \zeta \left(2q+2j\right) +  2 T(q,j) \zeta ' \left(2q+2j\right) \right\}.
\label{eq13}
\een
We take into account in the expansion in Eq. (\ref{eq13}) only the
$c^2$-dependent terms up to the 
second order, which is consistent with the spirit of the GL model. The terms $\phi ^{6}$, $\phi ^{8}$, ..., are considered as irrelevant in the neighborhood of the transition. So, Eq. (\ref{eq8}) becomes  
\ben
V(c, l)  & \approx & -  \frac{(-\pi) ^{p}}{p! 2 l^{2p+1}} 
 \left\{ -2c^{2p}\ln{c} - 2 c^2 \left[ \zeta (-2p+2)  \right. \right. \nonumber \\
&  - & \left.  2p \zeta '(-2p+2) \right]   +  c^4  \left[ (-2p+1) \zeta (-2p+4) \right.
\nonumber \\
& - & \left. \left. 2p (-p+1)\zeta '(-2p+4) \right] \right\}.  
\label{eq14}
\een
Hence, considering the particular case of our interest, $d=3$  ($p=1$),
and writing $c$ explicitly in terms of $\phi$ and $l$,  we obtain the correct
expression for the effective free energy density, 
\ben
{\cal F} (\phi; l) \approx   \frac{1}{2} m^2 \phi ^2 + \frac{u}{8 } \phi ^4 
- v \phi ^2 \ln{\phi} ,
\label{eq15}
\een
with the following definitions of the coefficients $m^2, \; u $ and $v$,
\ben
m^2 & = & m_0 ^2 +  v \left( 1 -  2 \ln{ e  l } \right), \label{eq16} \\
u & = & \lambda - \frac{e^4 l}{24 \pi }, \label{eq17} \\
v & = &  \frac{e^2}{4 \pi l }.
\label{eq18}
\een

Notice that the concavity condition for the free energy (\ref{eq15}) restricts the model to strictly positive values of $u$ . This implies, using Eq. (\ref{eq17}), in the upper limit for the film thickness,
\be
l^{max} = \frac{24 \pi \lambda }{e^4}.
\label{eq20}
\ee
It is also worthy mentioning that Eq. (\ref{eq15}) above has the same {\it formal} dependence on the order parameter $\phi$ as in Ref. \cite{FSU}. Eq. (\ref{eq15}) describes a first-order transition in systems in a form of a film. However, we have obtained a logarithmic $l$-dependence in Eq. (\ref{eq16}) above for the coefficient of the $\phi ^2$ term, as well as a linear dependence on $l$ for the self-coupling constant $u$, given by Eq. (\ref{eq17}).

In order to have a better understanding of the $l$-dependence of this phase transition, we can study the behaviour of the system from Eq. (\ref{eq15}), in a similar way as is done in Refs. \cite{HLM} and \cite{FSU}. Taking the first derivative of ${\cal F}$ with respect to $\phi$, we have ${\cal F}'  = \phi g$, where the function $g=g(\phi)$ is 
\be
g =  m^2 - v + \frac{u}{2 } \phi ^2 
- v \phi ^2 \ln{\phi}.
\label{eq19}
\ee
Therefore, for the equation of state ${\cal F}' = 0$, we get the following possible solutions: $\phi \equiv \phi_{0N} = 0 $, which describes the normal phase,  and positive solutions coming from $g=0$, which correspond the broken phase. We denote them by $\phi _{0B}$. Furthermore, the solutions must produce minima in ${\cal F}$ to be stable, i.e. they must obey the condition ${\cal F}'' (\phi _0)>0$. Then, since ${\cal F}''  = g' \phi + g$, we clearly see the normal phase is always stable. On the other hand, in the superconducting phase we have $ \phi _{0B} > 0$ and $g' (\phi_{0B}) >0$, which yields $ \phi_{0N} > \sqrt{\frac{2v}{u}} $. Notice that these solutions are real and finite for $u > 0$ (which as already remarked is also the condition of concavity of the model).
In addition, we can obtain a particular value of $ \phi _{0B}$ for which the free energy density vanishes. Writing ${\cal F}$ in terms of $g$, given by Eq. (\ref{eq19}), we obtain for ${\cal F}(\phi _{0B})= 0$ the value $\phi _{0B}  \equiv \phi _{0B,E} = 2\sqrt{\frac{v}{u}}$. Notice that there are different values, $\phi _{0N}$ and $\phi _{0B,E}$, which imply in ${\cal F}=0$; this situation corresponds to the equilibrium transition point of the first order transition, as remarked in Ref. \cite{FSU}.  

Thus, it is possible to obtain the value of the coefficient $m^2$ at the equilibrium point, ${\cal F}(\phi _{0B,E})= 0$; we get
\be
m^2 _E = v \left( \ln{\frac{4v}{u} } -1 \right).
\label{eq21}
\ee
So, from Eq. (\ref{eq16}) and expliciting the dependence of $u$ and $v$ in $l(=\xi_0 / L)$ we get the following expression for the $l$-dependent equilibrium transition temperature, 
\be
T_E (l)= T_0 \left[1 + \frac{e^2}{4 \pi l }\left( \ln{\frac{24 e^4 l}{24 \pi \lambda - e^4 l} } - 2 \right)\right].
\label{eq22}
\ee
To obtain an quantitative estimate from Eq. (\ref{eq22}), let us remember the 3-dimensional expressions for the coupling constants \cite{K}, 
\begin{equation}
\lambda \approx 111.08 \left(
\frac{T_{0}}{T_{F}}\right)^{2},\; e \approx 2.59
\sqrt{\frac{\alpha v_{F}}{c}}, 
\label{eq23}
\end{equation}
where $T_{F}$, $v_{F}$ and $\alpha$  are respectively the Fermi
temperature, Fermi velocity and fine structure constant (remember also $l=\xi_0 / L)$).

%%%%%%%%%%%%%%%%
\begin{figure}[ht]
\includegraphics[{height=8.0cm,width=8.0cm}]{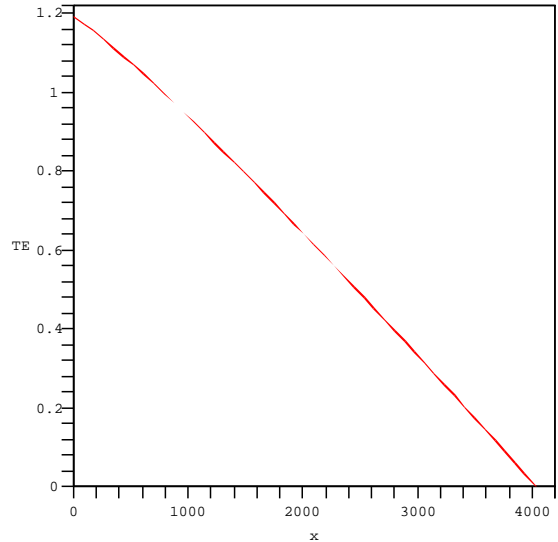}
\caption{ Plot of the equilibrium transition temperature $T_{E}$ for aluminium as function of $x= \frac{1}{l}$.  }
\label{fig}
\end{figure}
%%%%%%%%%%%%%%%%

Considering as example an aluminum sample, and taking their tabulated values, $T_{0}=1.19^{o}K$, $T_{F}=13.6 \times 10^{4}K$, $v_F = 2 \times 10^6 m/s $ and $\xi_{0}=1.6 \mu m$, in Fig. (\ref{fig}) is plotted $T_E (l)$ written in Eq. (\ref{eq22}) as function of $1/l$. It suggests that there is a minimum thickness below which there is no equilibrium transition temperature. For Al, it is $l_{min}^{Al}=1 / 4036.55$, and remembering that $l=L/ \xi_0$, we obtain $L_{min}^{Al} = 2.48 \times 10^{-4} \xi _0 $, i.e. $L_{min}^{Al} \approx 4 \; Angstrom$. Moreover, notice from Eq. (\ref{eq20}) that our model is restricted to films of thickness lower than $L_{max}^{Al}=12.4 \xi_0 \approx 18.4 \times 10 ^{-6}$ (in Fig. (\ref{fig}) this corresponds to a minimum value of $x$, $x_{min} \approx 1 / 12.4$). If we take in 
Eq. (\ref{eq14}) only the $l$-independent part of the $c^4$term we would obtain $u=\lambda$, 
$i.e$, no $l$-correction to the self-coupling constant. In this case, the condition 
$c^2 \ll 1$ would give a slightly higher upper limit, $L_{max}^{Al}\approx 20.1\xi_0$. 
In other words, taking into account the $l$-correction to the self-coupling constant constrains the film to be slightly thinner. Also it should be noticed that we have used 3-dimensional values for the coupling constants. This means that we should restrict ourselves only to relatively thin films; very thin films {\it can not} be physically accommodated in the 
context of our model. The value we have obtained for $L_{min}^{Al} (\approx 4 \; Angstrom) $ corresponds to 2-dimensional systems, and so, on physical grounds, is beyond the domain of validity of the model. 

Now, let us derive the expression of the size temperature of the first-order transition, defined by $\left( \Delta T \right)_{E}=\left| L(T_{E}) / \Delta C(T_{E}) \right|$, where $L(T_{E})$ is the latent heat at $T_E$ and $\Delta C(T_{E})$ is the specific-heat jump.  The latent heat is obtained from $L(T_{E})=T_{E}\Delta S(T_{E})$, where $\Delta S(T_{E})=S_{0N}(T_{E})-S_{0B}(T_{E})$ is the jump in the entropy ($S= d {\cal F} / d T$) at $T_E$ between the two phases. Also, specific-heat jump comes from $ \Delta C(T_{E}) = T_E \left( \frac{d S_{0N}(T_{E})}{d T} - \frac{S_{0B}(T_{E})}{d T} \right)$. Hence, the expression for the size temperature is 
\be
\left( \Delta T \right)_{E} =  T_0 v \equiv \frac{T_0 e^2 \xi_0}{4 \pi L}.
\label{eq24}
\ee
To obtain an numerical estimation, we can come back to the aluminium example and consider a sample of $L \approx 6 \xi_0 $, i.e. near the upper limit of validity of the model. We get $\left( \Delta T \right)_{E} ^{Al} = 6.1 \times 10^{-6} K$. We see that this value is greater about 3.5 times than that in \cite{HLM} for a bulk aluminium sample. It is important to remember that $\left( \Delta T \right)$ in \cite{HLM} is calculated with the specific-heat jump taken at the bulk second-order transition temperature $T_0 $, which differs of a factor 1/4 from $ \Delta C $ at the bulk first-order transition temperature, $T_{0E}$, as is pointed in Ref. \cite{ST}. Therefore, taking $ \left( \Delta T \right)^{Al} $ in \cite{HLM} calculated at $T_{0E}$ yields $ \left( \Delta T \right) ^{Al}_{0E} (HLM) = \left( \Delta T \right)^{Al}_{0} (HLM) / 4  \approx 1.75  \times 10^{-6} K $, which jutifies the claim of an enhancement in the size temperature of the transition for films.  

In conclusion, we have investigated the HLM effect in films using a 
compactified version of the 
GL model. As a consequence, the free energy (\ref{eq15}) has a logarithmic and linear 
dependence on the film thickness $L$ through the coefficients of the $\phi ^2$ and $\phi ^4$ 
terms, respectively. However, it should be noticed that  
in spite of these differences, the $L$-dependence of Eq. (\ref{eq24}) for the size 
temperature is formally the same as in Ref. \cite{FSU}. The size temperature of 
the transition has 
a $L^{-1}$-dependence, which means that the first-order phase transition is enhanced as $L$ decreases, 
in qualitative agreement with \cite{FSU}.
The equilibrium transition temperature of the first-order transition decreases with the 
lowering of the film thickness, implying formally the existence of a minimal film thickness 
$L_{min}$, 
below which the transition is suppressed. Also, we notice from Eq. (\ref{eq20}) that 
the values of $L$ have an upper bound, $L_{max}=24\pi \lambda \xi_0/e^4$.   
Thus the present model is restricted to films of thickness $L$ such that $L_{min}<L<L_{max}$.
For aluminium, we get respectively, $L_{max}^{Al}=12.4 \xi_0 \approx 18.4 \times 10 ^{-6}$ 
and $L_{min}^{Al} \approx 4 \; Angstrom$. However, as explained in the comments below 
Eq.(\ref{eq23}), we must restrict ourselves only to relatively thin films which could be considered 
as essentially 3-dimensional objects.
We {\it can not} on physical grounds, pretend to apply our formalism in its present form 
to very thin films, which are $two$-dimensional systems. In particular,  
the value we have obtained for $L_{min}^{Al} (\approx 4 \; Angstrom)$, is physically 
meaningless, far beyond the scope of our model. 

This work was partially supported by CAPES and CNPq (Brazilian
agencies). A.P.C.M.  is grateful for kind
hospitality at Centre de Physique Th{\'e}orique/Ecole
Polytechnique, where part of this work has been done.\\

\end{document}